\documentclass[a4paper,twocolumn,11pt,unpublished]{quantumarticle}
\pdfoutput=1
\usepackage[utf8]{inputenc}
\usepackage[english]{babel}
\usepackage[T1]{fontenc}
\usepackage{amsmath}
\usepackage{hyperref}
\usepackage{bm} 

\usepackage{tikz}
\usepackage{lipsum}

\begin{document}

\title{The problem of relaxation to equilibrium}

\author{S. Limandri}
\affiliation{Universidad Nacional de Córdoba, Facultad de Matemática, Astronomía, Física y Computación, M. Allende s/n, 5000 Córdoba, Argentina}
\affiliation{CONICET, Instituto de Física Enrique Gaviola (IFEG), 5000 Córdoba, Argentina}
\orcid{0009-0004-1547-8434}
\author{S. Segui}
\affiliation{CONICET, Instituto de Física Enrique Gaviola (IFEG), 5000 Córdoba, Argentina}
\orcid{0000-0001-5616-4766}
\author{B. Castellano}
\affiliation{Universidad Nacional de Córdoba, Facultad de Matemática, Astronomía, Física y Computación, M. Allende s/n, 5000 Córdoba, Argentina}
\author{I. Belitzky}
\affiliation{Universidad Nacional de Córdoba, Facultad de Matemática, Astronomía, Física y Computación, M. Allende s/n, 5000 Córdoba, Argentina}
\author{G. Castellano}
\affiliation{Universidad Nacional de Córdoba, Facultad de Matemática, Astronomía, Física y Computación, M. Allende s/n, 5000 Córdoba, Argentina}
\affiliation{CONICET, Instituto de Física Enrique Gaviola (IFEG), 5000 Córdoba, Argentina}
\orcid{0000-0003-0964-0875}
\email{gustavo.castellano@unc.edu.ar}
\maketitle

\begin{abstract}
When a thermodynamic system is released from any constraint, after some time its evolution will render it into an equilibrium state. Although the description of this relaxation to thermodynamic equilibrium has been attempted through both classical (Hamilton's equations) or quantum (Schrödinger equation) approaches, no success has been achieved without recurring to additional hypotheses. The present work demonstrates the possibility of reaching equilibrium states in a simple classical gaseous system, by imposing non-strict boundary conditions, in the sense that all interactions with the container walls must occur according to Heisenberg's uncertainty principle.
\end{abstract}



%
%
%

\section{Introduction\label{s:intro}}

Thermodynamic equilibrium states are defined as those for which the repeated measurement of a macroscopic property furnishes a stationary outcome. The existence of such states is accepted as a zeroth postulate for developing the thermodynamic theory \cite{callen,struchtrup}, and fortunately accompanies our everyday experience. Whenever a thermodynamic system is released from any constraint, after some time its evolution will drive it to an equilibrium state.

The description of this relaxation to thermodynamic equilibrium has been attempted through both classical or quantum approaches. If an $N$-particle system can be considered as classical ---for example, in an ideal gas regime---, its behaviour is governed by a Hamiltonian $H^N$ through the corresponding canonical equations of motion: the global probability density $\rho(\bm{X}^N)$ for each dynamical state $\bm{X}^N \equiv (\bm{q}^N ,\bm{p}^N)$ representing all generalized coordinates $\bm{q}^N$ and momenta $\bm{p}^N$ must obey Liouville's equation \cite{liouville1838}
\begin{equation} \label{liouville}
 \frac{\partial\rho}{\partial t} = -\left\{ H^N ,\rho \right\} = -i\hat{L}\,\rho \;,
\end{equation}
where $\left\{ \cdot ,\cdot \right\}$ stands for the Poisson bracket. Since in the vector space containing all the possible probability densities, the Liouville operator $\hat{L}$ is Hermitian, solutions will always bear oscillatory components: clearly, classical mechanics cannot predict relaxation to thermodynamic equilibrium.

This suggests that a quantum approach must be stated. For non-relativistic systems, the single particle Schr\"odinger equation
may be dealt with for time independent Hamiltonians $\hat{H}$ through variable separation, to furnish the well-known stationary equation
\[
 \hat{H}\phi = E\, \phi \;,
\]
allowing to find the eigenenergies $E_n$ for the corresponding stationary states $\phi_n(\bm{r})$, which constitute a basis for the Hilbert space containing all the possible wave functions $\psi(\bm{r},t)$. This is the conventional fashion in which simple or more complicated situations may be solved, as occurs with the many-particle systems involved in thermodynamics. The general solutions are obtained from an initial state $\psi_o(\bm{r})$ by applying the time propagator operator $\exp(-it\hat{H}/\hbar)$
\[
 \psi(\bm{r},t) 
  = e^{-\frac{it\hat{H}}{\hbar}} \psi_o(\bm{r}) \;.
\]

In a quantum description for a many-particle system to occupy the different possible states, the probabilities are represented by means of the density operator $\hat{\rho}$ \cite{vneumann1927}. The Schr\"odinger equation 
leads to the Liouville - von Neumann equation
\[
 \frac{\partial\hat{\rho}}{\partial t} = -\frac{i}{\hbar} [\hat{H},\hat{\rho}] = -i\hat{L}\, \hat{\rho} \;,
\]
where $[\hat{H},\hat{\rho}]$ is the commutator of the total quantum Hamiltonian operator $\hat{H}$ and $\hat{\rho}$, which allows to define the \textsl{quantum Liouville operator} $\hat{L}$, in this case acting on operators of the corresponding Hilbert space (for example, on $\hat{\rho}$). It can also be shown \cite{reichl16} that a description in this framework provides solutions involving oscillatory components
\begin{widetext}
 \begin{equation}
  \hat{\rho}(t) = \sum_{n,m} \langle E_n | \hat{\rho}(0) | E_m \rangle \, e^{-it\,(E_n -E_m)/\hbar} \, |E_n\rangle \langle E_m | \;,
 \end{equation}
\end{widetext}
\textit{i.e.}, no relaxation to equilibrium may be attained, even under a quantum mechanical description.

Many different assumptions have successively been introduced to overcome this problem, which range from chaotic hypotheses \cite{gallavotti1999} to non-Hermitian theories \cite{ruter2010,ashida2020}, coarse-grain descriptions \cite{wehrl78,safranek2019b,strasberg2021,jug2024} or coherence analysis in collective quantum states \cite{zurek1994,lebowitz1999,rigol2008}. The solutions quoted above, however, are ideal, in the sense that the boundary conditions imposed are always \textit{extremely} accurate. Of course, real systems must afford fabrication details, small nonuniformities, or natural irregularities, and at a microscopic level, the Heisenberg uncertainty principle must be complied with. For instance, systems involving quantum dots may be a few nanometres in size, and their electronic or optical properties strongly reflect their quantum nature \cite{kovalenko15,shishodia23},
as evidenced in quantum-dot lasers and masers \cite{liu15,mantovani19}.
The real evolution of a system should be expressed in terms of the basis $\varphi_n(\bm{r})$ associated to the \textit{instantaneous} boundary conditions, which would randomly change to obey the uncertainty principle and certainly influence the \textit{instantaneous} eigenenergies. The analytical solution for such a problem is very complex; however, for a system as simple as a particle in a box, an initial pure state is expected to ``blur'' in a combination of the elements of the \textit{current} basis, maintaining the expectation energy value, since $\hat{H}$ is independent of time. Instead, important changes should arise when describing many-particle systems, like those analysed in the framework of thermodynamics.

It is clear that all solutions provided by Schr\"odinger equation always refer to ideal boundary conditions, which maintain constant values along any evolution of the system described. However, Heisenberg's principle states that every physical magnitude bears some uncertainty, as occurs even with the classical velocities and positions of molecules in a gas contained within a vessel. Though no apparent influence might be expected, when a gas in a high temperature or low density regime evolves, where molecules may be conceived as classical masses obeying Hamilton's equation, every time they bounce on the recipient walls, the uncertainty principle imposes some (tiny) blurring in the returning positions, associated to a consequent spreading of their linear momenta. The present approach attempts to provide the means to achieve a decay to thermodynamic equilibrium by only recurring to the Heisenberg uncertainty principle \cite{heisenberg1927}. Along this work, the evolution of a very simple system like a one dimensional classical ideal gas will be numerically represented, in order to achieve thermodynamic equilibrium after a number of elastic rebounds on the walls.


\section{Classical non-interacting molecules in one dimension\label{s:model}}

Following the hypotheses for an ideal gas in equilibrium, a set of $N$ independent point molecules with mass $m$ is allowed to freely evolve within a 1D recipient, located in the interval $[-L/2,L/2]$. In order to investigate the possible relaxation of this system towards equilibrium, the particles start uniformly distributed in half vessel, in the interval $[0,L/2]$; this distribution can be expressed in terms of the Heaviside function $\Theta$ as
\begin{equation} \label{dist-eq}
 \frac{2}{L}\Big[\Theta(x)-\Theta(x-L/2)\Big] \;.
\end{equation}
An initial thermodynamic equilibrium state at temperature $T$ within this interval is furnished following Boltzmann distribution for their kinetic energies. Introducing the parameter $\beta=1/(k T)$, where $k$ is the Boltzmann constant, the one-particle probability distribution $\rho_1$ can be written as
\begin{widetext}
\[
 \rho_1(x,p;t\!=\!0) = \frac{1}{\sqrt{2\pi mk T}}\frac{2}{L}\Big[\Theta(x)-\Theta(x-L/2)\Big]\, e^{-\beta\frac{p^2}{2m}} \;.
\]
\end{widetext}
All the prefactors in this expression ensure an adequate normalization condition
\[
 \int_{-L/2}^{L/2}\text{d}x\int_{-\infty}^{\infty}\text{d}p\; \rho_1(x,p;0) = 1 \;.
\]
Pseudo random number generators allow to obtain the desired uniform distribution in coordinates, as well as the Gaussian momentum distribution, for example by means of the Box-M\"uller method \cite{rambo2013-BM}. By this procedure, each particle $i$ is assigned a linear momentum $p_i$, which in principle remains constant, since no energy losses are allowed when bouncing on the walls, following the ideal gas hypotheses for equilibrium \cite{kotz06}. As mentioned above, the gas will \textit{not} spread towards an equilibrium state under these assumptions, \textit{i.e.}, it will not occupy the whole interval $[-L/2,L/2]$ in a stationary state.

The general form for the classical Liouville's equation \eqref{liouville} with the restrictions imposed in this simple coordinate-bounded example relies on the fact that the corresponding (classical) Liouville's operator is Hermitian, as shown here. If $\eta$ and $\rho$ are two possible elements from the probability distribution vector space, denoting $\langle f|g\rangle$ as the scalar product of two functions $f\,$ and $g$ in this (complex) vector space, the proof for the Hermiticity of the classical Liouville operator follows directly by computing the scalar product
\begin{widetext}
\[
 \big\langle\eta\big|\hat{L}\rho\big\rangle = -i\,\big\langle\eta\big|\left\{H^N,\rho\right\}\big\rangle =
  -i\sum_j\int \mathrm{d}^N\!p\; \mathrm{d}^N\!q\; \eta^*
   \left(\frac{\partial H^N}{\partial p_j}\frac{\partial\rho}{\partial q_j} -
    \frac{\partial H^N}{\partial q_j}\frac{\partial\rho}{\partial p_j}\right) \;.
\]
\end{widetext}
The last member can be integrated by parts; considering Hamilton's equations, and taking into account that $\dot{p}_j$ becomes zero when approaching $p_j^\text{min}$ or $p_j^\text{max}$, and $\dot{q}_j=0$ when attaining $q_j^\text{min}$ or $q_j^\text{max}$,
\begin{widetext}
\begin{align*}
 \big\langle\eta\big|\hat{L}\rho\big\rangle = +i\sum_j&\Bigg\{\int \mathrm{d}^{N-1}p_{k\neq j}\;  \mathrm{d}^N\!q\; \bigg[\eta^* \underbrace{\frac{\partial H^N}{\partial q_j}}_{-\dot{p}_j} \rho\bigg]_{p_j^\text{min}}^{p_j^\text{max}} -\int \mathrm{d}^N\!p\; \mathrm{d}^N\!q\; \left( \frac{\partial H^N}{\partial q_j}
  \frac{\partial\eta^*}{\partial p_j} + \,
 \eta^*\frac{\partial^2 H^N}{\partial p_j\partial q_j}\right)\rho\, -\\
 & - \int \mathrm{d}^N\!p\;  \mathrm{d}^{N-1}q_{k\neq j}\; \bigg[\eta^* \underbrace{\frac{\partial H^N}{\partial p_j}}_{\dot{q}_j}\rho\bigg]_{q_j^\text{min}}^{q_j^\text{max}} 
  + \int \mathrm{d}^N\!p\; \mathrm{d}^N\!q\;  \left(\frac{\partial H^N}{\partial p_j} \frac{\partial\eta^*}{\partial q_j} +
\eta^*\frac{\partial^2 H^N}{\partial p_j\partial q_j}\right) \rho \Bigg\} \\
 \Rightarrow &\quad\big\langle\eta\big|\hat{L}\rho\big\rangle = \big\langle\hat{L}\eta\big|\rho\big\rangle \;.
\end{align*}
\end{widetext}


In order to verify these hypotheses, the evolution of the system is accomplished by allowing the point masses to move, maintaining their initial $p_i$ values constants after a short interval $\delta t$, \textit{i.e.}, a first attempt is
\[
 x_i(t+\delta t)=x_i(t)+\frac{p_i}{m}\,\delta t \quad \mbox{and} \quad p_i(t+\delta t)=p_i(t) \;.
\]
The $x$ values, however, must always remain within the interval $[-L/2,L/2]$, which imposes additional conditions. Depending on the values taken by $p_i$ and $\delta t$, the number of the particle $i$ bounces will be
\[
 n = \left\lfloor \left( x_i(t) + \frac{p_i}{m}\delta t + \text{sgn}(p_i) \frac{L}{2} \right) \bigg/ L \right\rfloor \;,
\]
where $\lfloor y\rfloor$ stands for the floor function (largest integer below the real variable $y$), which produces the correct ending value for position and momentum after the interval $\delta t$
\begin{align*}
 x_i(t+\delta t) &= (-1)^n \left[x_i(t) + \frac{p_i}{m}\delta t - nL \right] \;, \\
  p_i(t+\delta t) &= (-1)^n\, p_i(t) \;.
\end{align*}

The feasibility of this approach has been tested in a discrete representation of the intervals for coordinates and momenta, originally sampled as double precision variables. After several tests, an optimal discretization of $M=1000$ intervals for each variable ($x$ and $p$) was chosen. 
The system evolution has been surveyed in a system of $2^{27} \ (134,217,728)$ particles of $6.646474\times10^{-27}\,\text{kg}$ (4 neutron mass) within a $1\,\text{m}$ vessel, allowing $1\,\mu\text{s}$ steps. The code was written in C, and the computations were parallelized by means of the \textsc{omp} routines over $64$ threads @3.3\,GHz of a general-purpose \texttt{EPYC 7532} cluster (CCAD, Universidad Nacional de Córdoba, Argentina).

\begin{figure*}[t] 
\centering
\includegraphics[width=0.51\textwidth]{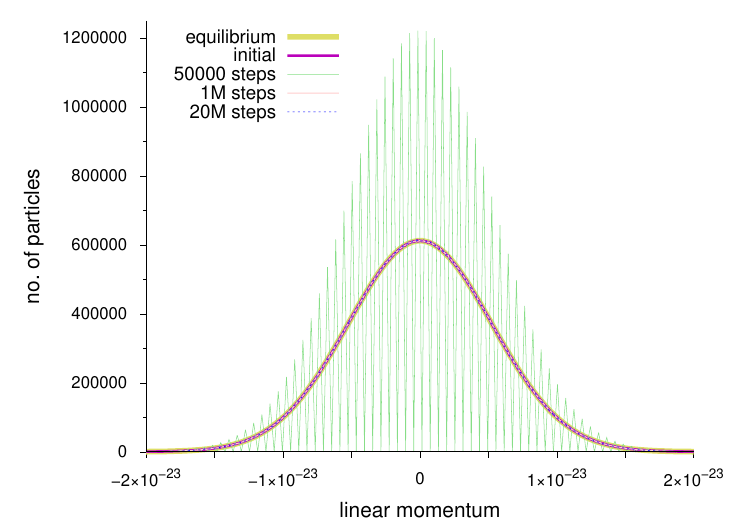}%
\includegraphics[width=0.51\textwidth]{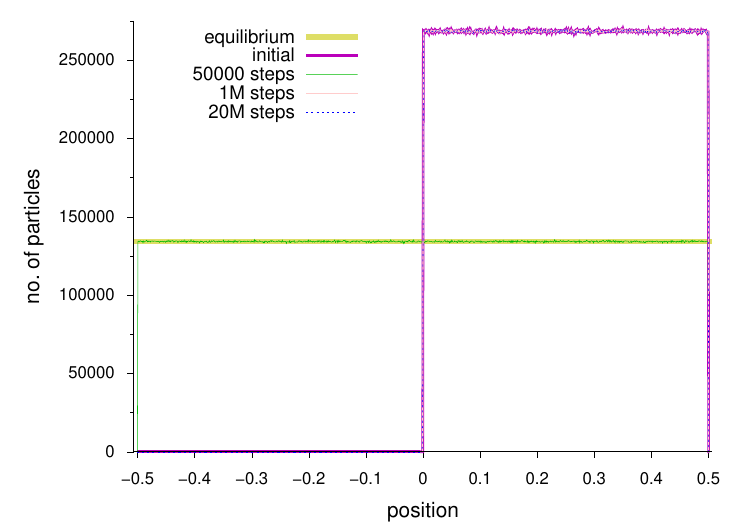}
\caption{Evolution of linear momentum (left) and coordinate (right) distributions for $2^{27} \ (134,217,728)$ particles out of thermal equilibrium, according to classical dynamics. Intermediate spreading for $50,000$ steps is also shown, and return to initial distribution after $10^6$ steps (1 complete cycle) and $2\times10^7$ steps (20 cycles), as predicted by the Liouville's equation.\label{nonoise}}
\end{figure*}

The periodic nature of this system under the constraints presented is better visualized if, in the \textit{initial} distribution, momenta are rounded to the centre value of each interval: this discretization permits reaching finite recurrence periods in reasonable computation times. These initial pseudo-random distributions have been generated by means of the 32-bit \texttt{xorshift} random number generator \cite{marsaglia2003}, according to the rules depicted above. For computational efficiency, the total number of particles has been split into 64 independent batches involving $2^{21} \ (2,097,152)$ particles, each of these groups taking less than one minute to run $10^6$ steps (1 complete period). After the total number of desired steps, all the resulting distributions were gathered by means of another program, also written in C.

Figure \ref{nonoise} displays the evolution obtained for the original Boltzmann distribution in half the recipient, which apparently spreads towards a spatial equilibrium delocalization, as can be seen in the configuration illustrated for $50,000$ steps---and reproduced in most intermediate states. For this particular case, the momentum distribution plot shows that some particles have been able to reach a wall, thus abandoning their initial momentum bin and therefore being transferred to the bin with opposite sign. This apparent reordering in linear momenta and spatial spreading in the recipient continues in all intermediate states, until the cyclic nature predicted by the Liouville's equation \eqref{liouville} is realised after an integer number of cycles ($10^6$ steps) has been completed. The double precision used for the individual dynamic variables ensures a stable reproduction for the system periodicity, as evidenced after $20$ cycles: as compared to the initial distributions, the slight distortion of the final ones is due to numerical rounding.

%

These results ensure that an appropriate description has been attained for this large set of free particles in a one dimensional recipient, which encourages to test how the restrictions imposed by the uncertainty principle may distort this periodic evolution for the simple system analysed. The following section shows that, despite the system sticks to the ideal gas assumptions, the allowance of \textit{slightly} changing the particle dynamic state only when bouncing on the walls, translates in the realization of relaxation to a thermodynamic equilibrium state.

\subsection{Introducing noise when bouncing on the walls\label{s:halfbox}}

\begin{figure*}[ht] 
\centering
\includegraphics[width=0.51\textwidth]{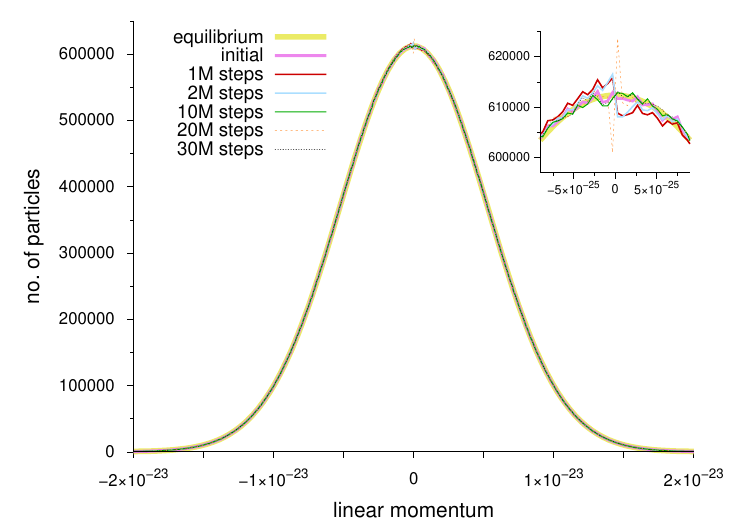}%
\includegraphics[width=0.51\textwidth]{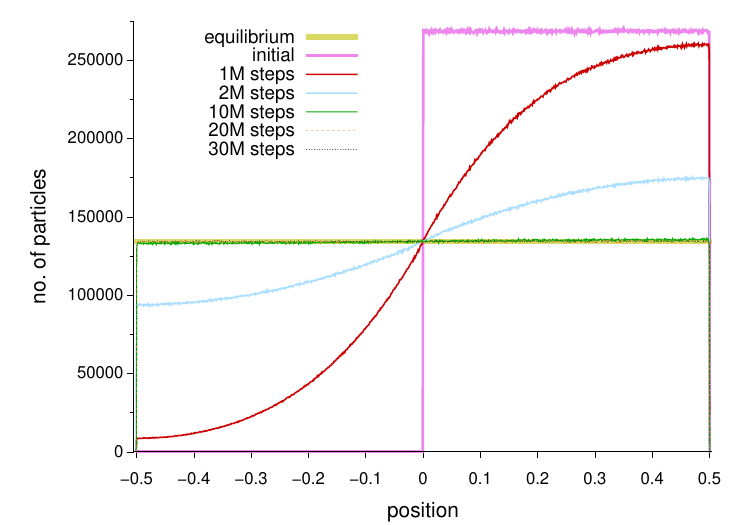}
\caption{Thermalization of the 1D ideal gas initially out of thermal equilibrium, consisting in $2^{27} \ (134,217,728)$ non-interacting particles. Linear momentum (left) and coordinate (right) distributions for different stages, imposing that every particle bounces at the walls obeying the Heisenberg uncertainty principle ---always following the ideal gas assumptions, excepting thermodynamic equilibrium. It is clear that the approach introduced avoids the cyclic nature of this evolution, as predicted by the Liouville's equation.\label{pm-mder}}
\end{figure*}

In order not to alter the ideal-gas basic hypotheses, the Heisenberg uncertainty principle has been introduced as follows. Every time a particle bounces on the borders ($x=\pm L/2$), the final position is randomly ``blurred'' by means of a Gaussian distribution with standard deviation 0.1\,mm. In addition, since its linear momentum also varies, its kinetic energy $\varepsilon_i$ is slightly changed by an amount $\Delta\varepsilon_i$ by means of a uniform random distribution that obeys $\langle\Delta\varepsilon_i\rangle=0$. This criterion has been assumed taking into account that these slight changes in the magnitude of each individual linear momentum $p_i$ cannot follow the same rules for all particles, since they are unable to deliver an amount $\Delta\varepsilon_i$ exceeding each value $\varepsilon_i$. The assumption $\langle\Delta\varepsilon_i\rangle=0$ ensures that at every step $\Delta t$, on average, the total energy of the system is conserved. The normalized probability distribution $f(\Delta\varepsilon_i)$ for the changes in $\varepsilon_i$ when bouncing on the walls was chosen as
\[
 f(\Delta\varepsilon_i) = \frac{1}{2\Delta_i}
  \big[ \Theta(\varepsilon_i-\Delta_i)+\Theta(\varepsilon_i+\Delta_i) \big] \;,
\]
where $\Delta_i\equiv\alpha\,(p_{\text{max}}-|p_i|)(|p_i|-p_{\text{min}})$; the magnitude of the constant $\alpha$ can be adjusted to increase the speed or slow down the run, always obeying the uncertainty principle, and also taking care not to to exceed the interval $[p_\text{min},p_{\text{max}}]$ for the allowed values of $|p_i|$. After several attempts, the probability density $f(\Delta\varepsilon_i)$ has been defined by choosing the parameter $\alpha=10^{-4}$.

It is worth mentioning that, since some uncertainty has also been allowed to the final particle positions after bouncing on the walls, they may occupy bins shortly outside the original $x$ interval $[-L/2,L/2]$. This translates as a couple of extra bins at each side, which means that the spatial histograms embrace positions between $[-L(1/2+2/1000),L(1/2+2/1000)]$, \textit{i.e.}, 1004 bins.

The approach presented in this section was also implemented in the general-purpose \texttt{EPYC 7532} cluster (CCAD, Universidad Nacional de Córdoba, Argentina) mentioned above, maintaining the initial conditions chosen for the previous example. In order to measure to which extent the attained distributions look alike the equilibrium configuration, the global measure of the goodness of these populations has been implemented through $\chi^2$-tests \cite{montgomery2019}, which are assessed as desired after a block of steps by comparing the sets of $M$ numbers of particles $\Delta N_j$ contained in each $j$ bin with the expected numbers $\Delta E_j$ associated to the thermodynamic equilibrium distribution \eqref{dist-eq}
\[
 \chi^2=\frac{1}{M-1}\sum_{j=1}^M\frac{(\Delta N_j-\Delta E_j)^2}{\Delta E_j} \;.
\]
These tests have been performed to simultaneously check the distributions for $x$ and $p$. It is worth mentioning that, in order to deal with populations statistically sensible, bins bearing less than $\simeq100$ counts were disregarded in the $\chi^2$-tests carried out ---which is particularly relevant in the case of the momentum distributions.

Figure \ref{pm-mder} shows the $p$ and $x$ resulting distribution after integral multiples of one period ($10^6$ steps). It is clear that at the end of the first cycle, the sharp spatial distribution in the right-hand side of the recipient has been blurred mainly because of the slight changes occurred in the particle linear momenta after each bounce---the influence of the uncertainty in their positions is almost imperceptible.

\begin{figure}[th] 
\centering
\includegraphics[width=\columnwidth]{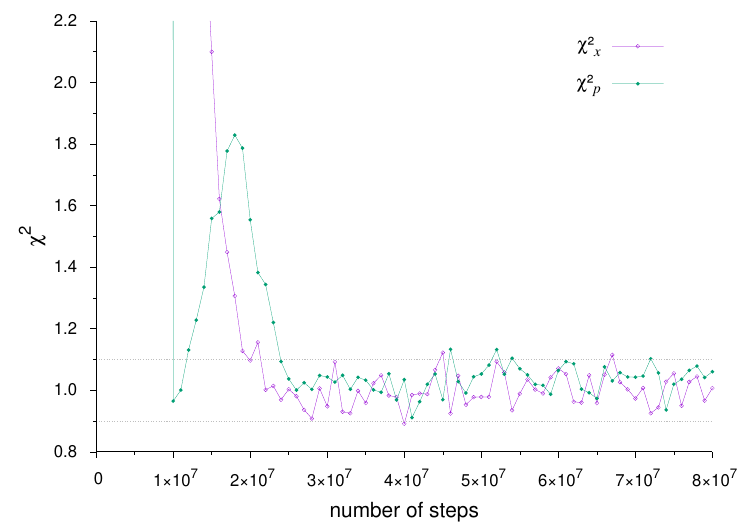}
\caption{Relaxation to thermodynamic equilibrium for a non-interacting 1D gas system beginning in the right-half of the recipient. The $\chi^2$ tests have been performed for distributions involving $2^{27} \ (134,217,728)$ particles, introducing the Heisenberg uncertainty principle when bouncing on the walls during the system evolution, as detailed in the text.\label{chi2-mder}}
\end{figure}

Figure \ref{chi2-mder} displays the behaviour of the $\chi_x^2$ and $\chi_p^2$ values, respectively corresponding to the coordinate and momentum distributions. Although in Fig.\ \ref{pm-mder} an apparent relaxation is achieved ``soon'' ($\approx10$ cycles), a careful analysis evidences the influence of the slow particles in the final distributions, since it is necessary to gain statistics after they also reach a reasonable number of bounces on the walls.

It can be seen that, with the settings depicted above, after a number of damped oscillations in $\chi^2_p$ (not displayed in Fig.\ \ref{chi2-mder}), in the present case around $30$ million steps were necessary to observe the expected decay to the equilibrium state. In other words, after 30 complete cycles the initial-condition memory for this system was completely ``erased'', and afterwards any successive distribution coincides with that corresponding to thermodynamic equilibrium. It is worth mentioning that for this extremely simple approach, without interactions between molecules of this 1D gas, a relaxation time of $30\,\text{s}$ ($=30\times10^6\times1\mu\text{s}$) is more than reasonable for the $1\,\text{m}$ container of the present example. Once the equilibrium state has been attained, the ideal gas original hypotheses can be recovered, and there is no need to continue with the `blurring' of variables in each collision with the walls.

\subsection{1D ideal gas moving towards right\label{s:other}}

\begin{figure*}[th] 
\centering
\includegraphics[width=0.51\textwidth]{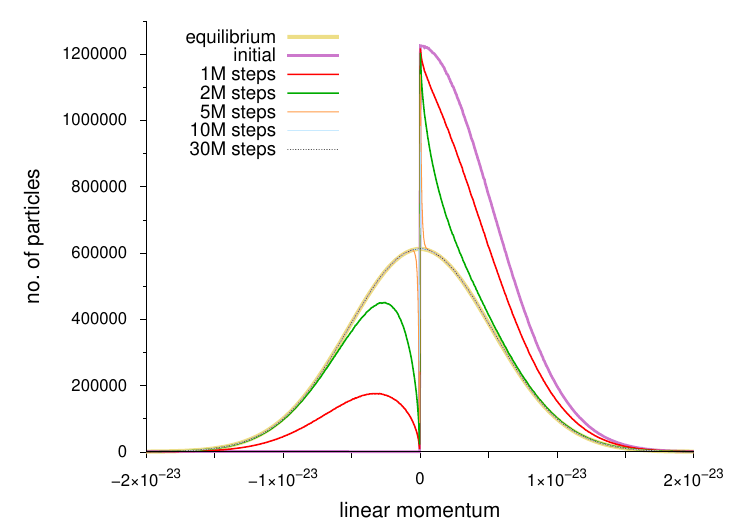}
\includegraphics[width=0.51\textwidth]{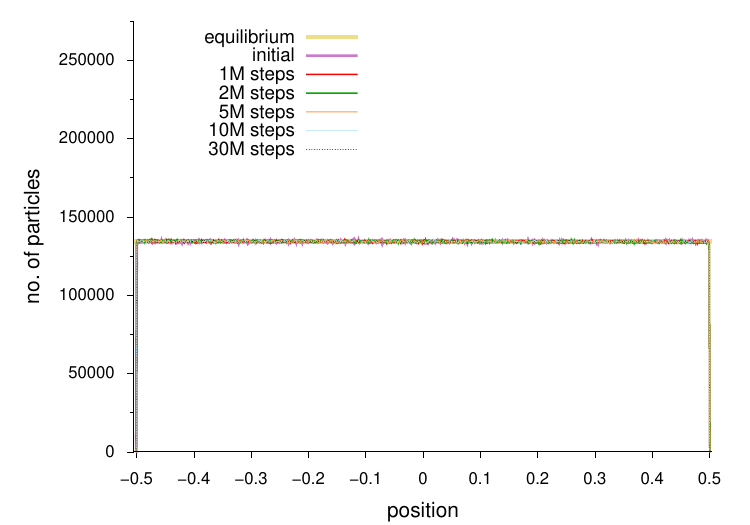}
\caption{Thermalization of the 1D ideal gas with initial velocities according to Boltzmann's distribution, allowing only positive values (all particles moving towards right). This 1D gas consists of $2^{27} \ (134,217,728)$ non-interacting particles, for which every bouncing on the walls smears their classical dynamical state, according to the uncertainties imposed by the Heisenberg uncertainty principle. At every step the ideal gas assumptions are obeyed ---excepting thermodynamic equilibrium. As can be seen, the present approach allows thermodynamic relaxation, despite the simplicity of the description.\label{px-p+}}
\end{figure*}

The present approach has also been tested for other simple cases. In the example shown in this subsection, the particles are initially assumed to be uniformly distributed in the vessel, and moving towards the right. The evolution of the system obeys the same rules as the previous case, \textit{i.e.}, the ideal gas original hypotheses, and allowing small uncertainties in the linear momenta and positions after bouncing in the recipient walls. Figure \ref{px-p+} displays the probability distributions for this situation at different stages, again evidencing that thermalization is achieved approximately after 30 periods, an interval similar to that required in the previous example.

It is interesting to notice that although the global linear momentum of the system is initially different from zero, the probability distribution $f(\varepsilon)$ for the changes in the individual particle energies guarantees that the total energy is conserved, whereas the final momentum vanishes---being indeed absorbed by the vessel, rigidly attached to the ground. 

\begin{figure}[th] 
\centering
\includegraphics[width=\columnwidth]{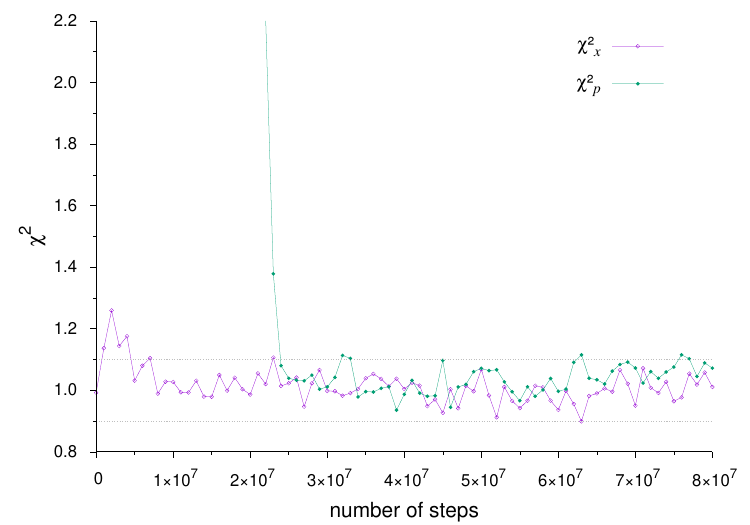}
\caption{Relaxation to thermodynamic equilibrium for a non-interacting 1D gas for which the initial condition is uniform spatial distribution within the vessel, moving towards the right wall ($p_i>0\;\forall\,i$, Boltzmann distribution). The $\chi^2$ tests have been performed for distributions involving $2^{27} \ (134,217,728)$ particles, as detailed in the text.\label{chi2-p+}}
\end{figure}

Figure \ref{chi2-p+} displays the evolution of the $\chi_x^2$ and $\chi_p^2$ values, providing tests similar to those shown above. Again, the influence of the slow particles in the final distributions is slower than the contribution corresponding to average kinetic energies, requiring a number of bounces on the walls to achieve reasonable statistics in the distributions.



In all the examples shown here, the uncertainty in the boundary conditions when bouncing on the walls can be interpreted in different ways. An alternative is to visualize that if the length of the recipient is not perfectly defined, it can be thought of at some coordinate for an arriving particle, and at a different position for the following ones: this artificial coordinate displacement $\Delta x$ may be associated with an uncertainty $\Delta p$, representing a gain or loss in the particle linear momentum. 

\section{Conclusion}\label{conclu}

The decay of a simple system to thermodynamic equilibrium has successfully been achieved, recurring only to fundamental physical laws, and thus avoiding chaotic hypotheses, coarse-graining assumptions, or non-Hermitian dynamic operators. Although the chosen example involves a one-dimensional ideal gas under a classical regime, it is clear that its evolution avoids oscillatory components due to the inclusion of small uncertainties allowed after each interaction with the recipient walls, obeying Heisenberg's principle.

This approach overcomes the limitation provided by the Liouville's equation for the thermalization of any classical system. The present study will soon be complemented with the application of a similar approach to a simple quantum spin system, which cannot reach the thermodynamic equilibrium without additional hypotheses, as predicted by the Liouville - von Neumann equation.

Although these are very simple systems, it must be emphasized that the Heisenberg uncertainty principle can be incorporated to the boundary conditions associated to any physical system, of course with increasing complexity.

\section*{Acknowledgements}

This work used computational resources from CCDA - Universidad Nacional de Córdoba (https://ccad.unc.edu.ar/), which are part of SNCAD MinCyT, Argentina. The authors acknowledge financial support from the \textit{Secretar\'{\i}a de Ciencia y T\'ecnica (Universidad Nacional de C\'ordoba)}, grant number 33620230101137CB.

%
%
%
%

\bibliographystyle{quantum}
\bibliography{r-1D}

\end{document}